# Electronic in-plane symmetry breaking at field-tuned quantum criticality in CeRhIn$_5$

F. Ronning[1], T. Helm[2], K. Shirer[2], M. Bachmann[2], L. Balicas[3], M. Chan[4], B. Ramshaw[4], R. McDonald[4], F. Balakirev[4], E. Bauer[1], and P.J.W. Moll[2,*]

[1]Los Alamos National Laboratory, Los Alamos, NM 87545, USA
[2]Max-Planck-Institut for Chemical Physics of Solids, 01187 Dresden, Germany
[3]National High Magnetic Field Laboratory, Florida State University, Tallahassee, FL 32310, USA
[4]National High Magnetic Field Laboratory, LANL, Mississippi-E536, Los Alamos, NM 87545, USA

*Corresponding address: Philip.moll@cpfs.mpg.de

Electronic nematics are exotic states of matter where electronic interactions break a rotational symmetry of the underlying lattice, in analogy to the directional alignment without translational order in nematic liquid crystals. Intriguingly such phases appear in the copper- and iron-based superconductors, and their role in establishing high-temperature superconductivity remains an open question. Nematicity may take an active part, cooperating or competing with superconductivity, or may appear accidentally in such systems. Here we present experimental evidence for a phase of nematic character in the heavy fermion superconductor CeRhIn$_5$. We observe a field-induced breaking of the electronic tetragonal symmetry of in the vicinity of an antiferromagnetic (AFM) quantum phase transition at H$_c$~50T. This phase appears in out-of-plane fields of H*~28T and is characterized by substantial in-plane resistivity anisotropy. The anisotropy can be aligned by a small in-plane field component, with no apparent connection to the underlying crystal structure. Furthermore no anomalies are observed in the magnetic torque, suggesting the absence of metamagnetic transitions in this field range. These observations are indicative of an electronic nematic character of the high field state in CeRhIn$_5$. The appearance of nematic behavior in a phenotypical heavy fermion superconductor highlights the interrelation of nematicity and unconventional superconductivity, suggesting nematicity to be a commonality in such materials.

Heavy fermion compounds are metals where localized 4f/5f orbitals hybridize with the delocalized conduction electrons resulting in the formation of heavy quasiparticles close to the Fermi level. In these materials the hybridization energy, the Kondo energy describing spin-interactions with the conduction electrons, and the Ruderman–Kittel–Kasuya–Yosida (RKKY) interaction energy mediating magnetic exchange between the f-electrons via the conduction band are often comparable and thus compete to define the ground state of a material. This subtle energetic balance can be readily tilted to favor a different state by moderate perturbations, such as chemical substitution, pressure or even magnetic fields. This multitude of distinct but energetically close-by ground states renders this class of materials an ideal setting to study quantum critical phase transitions between them.

Here we focus on the high field state of CeRhIn$_5$, which at ambient pressure and zero field hosts an anti-ferromagnetic order (AFM) of nominally localized 4f electrons at T$_N$=3.8K(*1*). However, the notion of fully localized moments fails to capture the enhanced electronic specific heat ($\gamma \sim 70 mJ/mol\, K^2$)(*2*) which suggests that a small fraction of f-character persists at the Fermi surface even in the ordered state. Hydrostatic pressure tends to increase the f-orbital hybridization and thus generally leads to delocalization into heavy f-hybridized bands. This scenario seems to apply well to CeRhIn$_5$ where a critical pressure p$_c$=23kbar marks a quantum critical point separating the low-pressure, local moment anti-ferromagnetic phase from the high-pressure, delocalized paramagnetic phase at zero temperature. This can be directly observed by the pressure induced change from a small to a large Fermi surface seen by quantum oscillation experiments, accompanied by a divergence of the quasiparticle mass as a direct result of the quantum critical point (QCP) (*3*). A dome of unconventional superconductivity with a maximal T$_c$ of 2.6K arises around the QCP and its AFM quantum fluctuations are a promising candidate for a pairing mechanism (*1*, *4*). $^{115}$In-Nuclear Quadrupole Resonance under pressure(*5*) and field-angle dependent specific heat measurements(6) provide direct experimental evidence for an unconventional d-wave superconducting order parameter in $CeRhIn_5$.

Alternatively, at ambient pressure the AFM order can also be suppressed by magnetic fields at a critical field $H_c$=50T, which is remarkably isotropic despite the anisotropy of the magnetic susceptibility at high fields(7). Unlike pressure, magnetic fields tend to localize the 4f electrons and thus the physical situation at the phase boundary is expected to be different in these two situations. Yet the phenomenology in high magnetic fields and under pressure shares nonetheless unexpected similarities. Similar to the dome of superconductivity, recently a correlated phase was discovered in fields larger than H*~28T as a dome connected to the field-induced quantum critical point at $H_c$(8, 9).

Our main experimental observation is the appearance of strong in-plane resistivity anisotropy in this high field phase. Crystal symmetry enforces an isotropic resistivity in the basal plane of a tetragonal system, and resistivity anisotropy thus conversely indicates a lowered symmetry of the electronic system. Reliable, direction-dependent measurements of resistivity in high magnetic fields are commonly difficult due to the high conductivity of high quality crystalline $CeRhIn_5$. To address this experimental issue, we have fabricated microstructures from a single crystal of $CeRhIn_5$ using Focused Ion Beam machining (details on the fabrication procedure were published in (9) and are further described in the supplement). Typical devices feature rectangular bars in the plane of $CeRhIn_5$ (Fig.1). The transport devices were aligned along different crystal directions to probe for preferred directionality of the symmetry breaking. The first design type probes resistivity at 45° in the plane along the [110] and [1-10] directions (Fig.1a), while the second is made to measure along the [100] and [010] directions. The zero field resistances of all devices are in quantitative agreement, excluding experimental artefacts due to strain or current path misorientation in the structures. The resistance at room temperature is 31.3µΩ cm and 0.3µΩ cm at 2K, and the large residual resistivity ratio in excess of 100 indicates the high material quality of the crystal after the FIB process. The temperature dependence of the resistivity including fine details such as the s-shape around 200K and the sharp drop around 30K are in excellent agreement with previously published measurements of macroscopic bulk crystals of $CeRhIn_5$(10). Thus the microstructured devices shows no indication of damage or alteration due to the fabrication process, which is further supported by the large quantum oscillations readily observed in them at low temperatures(9).

Figure 2 encompasses the main experimental finding for both design types probing orthogonal resistance bars along [100],[010] and [110],[1-10] as shown in Figure 1. The magnetoresistance of each device shows a resistive anomaly for fields around H*, signaling the entry into the high-field phase. At the same time, a substantial asymmetry between the orthogonal in-plane resistance bars (red, blue) signals the broken symmetry in the plane. While one direction shows a strong increase in resistance, in the other the resistance decreases. A small applied in-plane field is key to determining the direction of low and high resistance. Here, the field is applied at a 20° angle off the c-axis, and tilted along one of the resistance bars (left red, right blue). The direction of low resistance in all studied samples was consistently observed in the resistance bar parallel to the in-plane field component.

The situation is remarkably similar for fields along the [100] and [110] directions, as shown in the rows of Figure 2. The phenomenology as well as the magnitude of the resistivity anisotropy appears to be largely independent of the underlying crystal lattice directions. Importantly, the field scale of H* as well as the magnitude of the resistivity anisotropy is identical for both field directions. The freedom to turn the anisotropic phase in the plane is suggestive of an XY-like character of the high field order, instead of an Ising-like behavior. As the field is tilted closer to the c-axis, the anisotropy decreases and consistently follows the angle dependence of the resistance jump previously reported in ref(9). When the field is applied as parallel to the c-direction as experimentally possible in pulsed fields (+/-2deg), the behavior is found to be device dependent (See supplement). In the absence of an in-plane field aligning the nematic direction, secondary effects such as strain distributions, local defect densities and the shape anisotropy appear to induce a domain pattern within the microstructure.

To simplify the comparison between geometries and samples, we define an anisotropy as the ratio of the resistivity along the in-plane field and the resistivity perpendicular to it, $\rho_\parallel/\rho_\perp$. For fields 20° off the c-axis, the in-plane anisotropy exceeds a factor of 5 at H* and successively decreases at higher magnetic fields until it vanishes into an isotropic state at the AFM phase boundary (Fig.3). When $H_{ip}$ is rotated from one leg of the device into the other, the resistivity anisotropy is reversed and perfectly mirrored. At the same time, a small anisotropy is observed even at fields below H*. This small anisotropy is a result of the difference between the longitudinal ($I \parallel H_{ip}$) and transverse ($I \perp H_{ip}$) magnetoresistance in the presence of the small in-plane field, which vanishes with $H_{ip}$ as the field is turned closer to the c-axis as expected (See supplement). The onset field is almost temperature independent describing a vertical line of phase transitions, similar to the pressure induced transition into the superconducting state(4) (Fig.3b). With increasing temperature, the anisotropy gradually shrinks until no anomaly at H* is observed at temperatures above T*~2.2K. At higher fields the anisotropy gradually vanishes, and this upper field scale traces well the field dependence of the Neel-line $T_N(H)$.

A remarkable feature of the transport anisotropy is that it develops exclusively in the plane. While the in-plane resistance is characterized by a large jump, the resistance along the c-direction remains featureless (See Fig2c and ref(9)). Despite the presence of strongly three-dimensional electronic bands in $CeRhIn_5$(11, 12), the in-plane resistivity anisotropy appears to occur in a highly two-dimensional electronic system, which does not influence electronic transport along the c-direction. At the same time, the absence of magnetic anomalies in the field range of 20-50T suggests that metamagnetism and associated anisotropic spin-scattering is not a likely explanation for the resistivity anisotropy. To search for magnetic transitions under the field configuration where the anisotropy is most pronounced ($\theta = 20°$), the magnetic torque was measured on the same macroscopic crystal from which the microstructures were fabricated. The torque crosses smoothly through the critical field H* without any sign for a magnetic transition (Fig.2b), consistent with previous pulsed magnetic field experiments showing a quasi-linear increase of the magnetization up to the AFM phase boundary(7). The absence of high field metamagnetism may be understood from the known low-field magnetic structure. In zero field, the AFM state consists of anti-ferromagnetically aligned planar spins on the Ce site forming an incommensurate spin-spiral along c with an ordering vector q=(1/2,1/2,0.297)(13). At relatively low in-plane fields of 2T, the spiral alignment of the spins along the c-axis becomes energetically unfavorable and the moments align perpendicular to the in-plane component in a metamagnetic spin reorientation, locking into a commensurate order of q=(1/2,1/2,1/4)(14). This metamagnetic transition is observed as a sharp anomaly in the magnetization and specific heat(15). Consistently, the reorientation is also evident as a small step in the resistivity at a field of 5T, which corresponds to 2T of in-plane field at 20° as expected from its $\sin(\theta)^{-1}$ angle dependence(16) (Fig.2). In fields beyond the metamagnetic transition, a gradual canting of the spins towards the eventual field-polarized paramagnetic state at 50T has been proposed and found to explain the high-field magnetic behavior well(7). Thus no present experimental evidence points towards metamagnetic transitions in intermediate fields.

An alternative scenario involves the in-plane symmetry breaking by itinerant charge carriers in an electronic nematic state(17). In analogy to liquid crystal phases in fluids that break rotational while conserving translational symmetries, the electronic nematic describes a correlated state of matter that breaks a rotational symmetry of the underlying Hamiltonian while preserving translational symmetries. Electronic nematic phases are an elusive state of matter, with only a few candidate systems such as the two-dimensional metal $Sr_3Ru_2O_7$, (Al,Ga)As heterostructures or high-temperature superconductors. Indeed the observed phenomenology is strikingly reminiscent of the field induced transition in $Sr_3Ru_2O_7$. A key experiment supporting a nematic state in $Sr_3Ru_2O_7$ is a pronounced field-induced in-plane resistivity anisotropy that is incompatible with its tetragonal crystal symmetry. While fields along the c-direction induce the transition, a small in-plane field component is found to select the in-plane direction of high resistance. The temperature dependence and functional form of the resistivity anisotropy(18) is strikingly similar to our results in $CeRhIn_5$. Yet while in $Sr_3Ru_2O_7$ an increase of the resistivity is observed in both orthogonal in-plane directions, in $CeRhIn_5$ the resistance increases in one and decreases in the other direction. This behavior is reminiscent of observations in other nematic candidates such as AlAs-GaAs heterostructures(*19*).

Importantly, "true" electronic nematicity is an idealized abstract concept as nematicity in real materials always appears intertwined with other order parameters. Symmetry dictates that once the rotational symmetry is broken by one order, any other order parameter belonging to the same broken symmetry is allowed to appear. For example, recent neutron diffraction experiments in $Sr_3Ru_2O_7$ found evidence for a spin-density wave (SDW) at finite momentum (*20*). Thus in strongly interacting electron systems, electronic nematicity in tetragonal crystals is often inseparably entangled with orthorhombic lattice distortions or spin-textures, such as the iron-based superconductors(*21*). This may also be the case in $CeRhIn_5$. A careful investigation of the high field magnetization (Fig.7 in ref.(*7*)) may indicate a minute change of the magnetic susceptibility in fields around 30T, which may be a signature of the spin texture relaxing into the new symmetry of the electronic system.

In recent years, the intimate connection between electronic nematicity and unconventional superconductivity has attracted significant attention. In particular in the iron-based superconductors, electronic nematic fluctuations have been identified to diverge at the doping-driven quantum critical point(*22*) and were proposed to be intimately connected with the highest transition temperature(*23–26*). In this case, aligning the nematic domains is best achieved via in-plane strain fields, which induce a large in-plane resistivity anisotropy(*27*). Now $CeRhIn_5$ is a particularly intriguing nematic candidate as both unconventional superconductivity in a heavy fermion material and the nematic state can be accessed in a stoichiometric compound by the cleanest possible tuning parameters, pressure and magnetic field. A direct comparison of the destruction of the AFM order by pressure and field indeed reveals unexpected parallels between the superconducting and the nematic state. Figure 4 contrasts the AFM phase boundary under pressure and in magnetic field, plotted against the dimensionless tuning parameter $g/g_c = H/H_c$ or $p/p_c$ respectively. The nematic phase bears remarkable similarities to the superconductivity observed under pressure: (1) both are bounded by a vertical line of first-order phase transitions; (2) they share the same temperature scale; and (3) both exist around the field- or pressure-driven destruction of the AFM order at similar values of $g/g_c$.

In light of these similarities, a common microscopic mechanism would be an interesting scenario unifying the experimental observations as sketched in Fig.4b. Crossing the AFM phase boundary at low fields under pressure induces d-wave superconductivity, while crossing it in large magnetic fields induces the electronic nematic. In both cases, the nematicity and the d-wave superconductivity are characterized by an electronically broken C4 rotational symmetry. With $CeRhIn_5$ identified as a candidate for an electronic nematic in the heavy fermion materials class, the connection between the high field phase and superconductivity can be directly explored, and mapping the field-pressure phase diagram at low temperatures will be critical to understand the interplay of these phases. Suppression of the AFM order by chemical substitution to zero temperature has been shown for example in the $CeRh(In,Sn)_5$ system(*28*), which may present a further route to reduce the field scales required to induce a nematic state to lower fields readily available in superconducting magnets.

An intriguing aspect of this apparent nematicity in $CeRhIn_5$ is the role of low electronic dimensionality. All prior candidate systems for electronic nematic materials such as $Sr_3Ru_2O_7$, cuprates (*29*) and iron-based superconductors and AlAs-GaAs heterostructures (*30*) are characterized by a strongly two-dimensional electronic structure. By contrast, $CeRhIn_5$ is a strongly three-dimensional metal as evidenced by its low resistivity anisotropy of 1.8 (Fig.2c) and quantum oscillation experiments(*12*). Yet despite this three-dimensional structure, the nematic phase transition appears to occur in a strongly two-dimensional electronic sub-system that dominates the in-plane transport but contributes negligibly to the out-of-plane conduction. Thus it again uncovers links between the nematic phase and superconductivity in $CeRhIn_5$, where detailed resistance measurements under pressure also indicate a strongly two dimensional character of the superconducting order(*31*). Together, these findings highlight once more the importance of low-dimensionality for electronic nematic phenomena and their entanglement with unconventional superconductivity, even where it is not expected such as in strongly three-dimensional metals like CeRhIn$_5$.

# Figures

## Figure 1: CeRhIn$_5$ microstructures for in-plane anisotropy

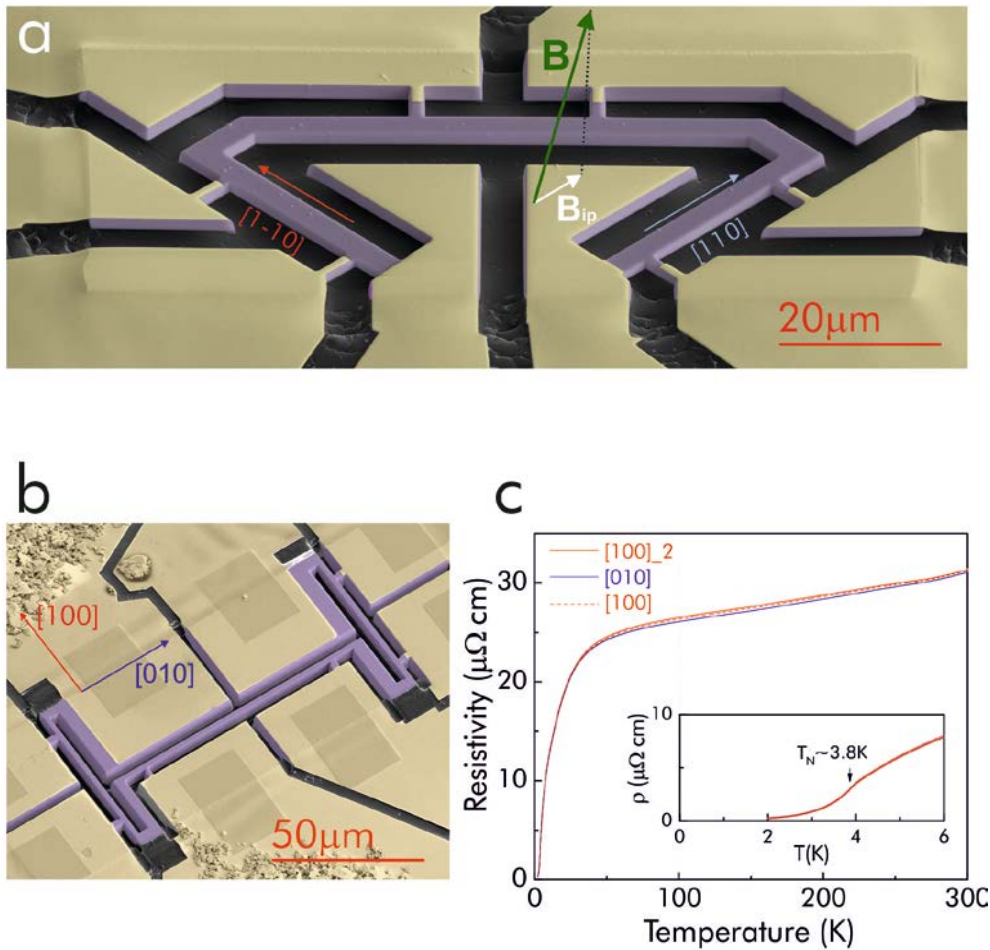

a) Electron beam micrograph of a CeRhIn$_5$ crystalline microstructure. The crystal lamella (purple) was carefully aligned so that the four-point resistance bars are along the [110], [1-10] and [100],[010] directions respectively. A current is injected through the bottom contacts and passes the entire U-shaped structure with all three bars in series. The dimensions of all bars were made exactly identical, with a length of 20µm, width of 4µm and thickness of 1.8µm. The field configuration indicating the small in-plane field is indicated as a sketch.

b) Related design to the microstructure shown in a). Here the transport bars were aligned with the crystal axes in the basal plane, [100] and [010].

c) Resistance of the devices as a function of temperature. All legs yield the same resistivity when scaled by the bar dimensions measured via SEM, indicating that the device is well aligned in the plane and that indeed no artificial in-plane asymmetry is present when none is expected.

Figure 2: Broken tetragonal symmetry in the high field state

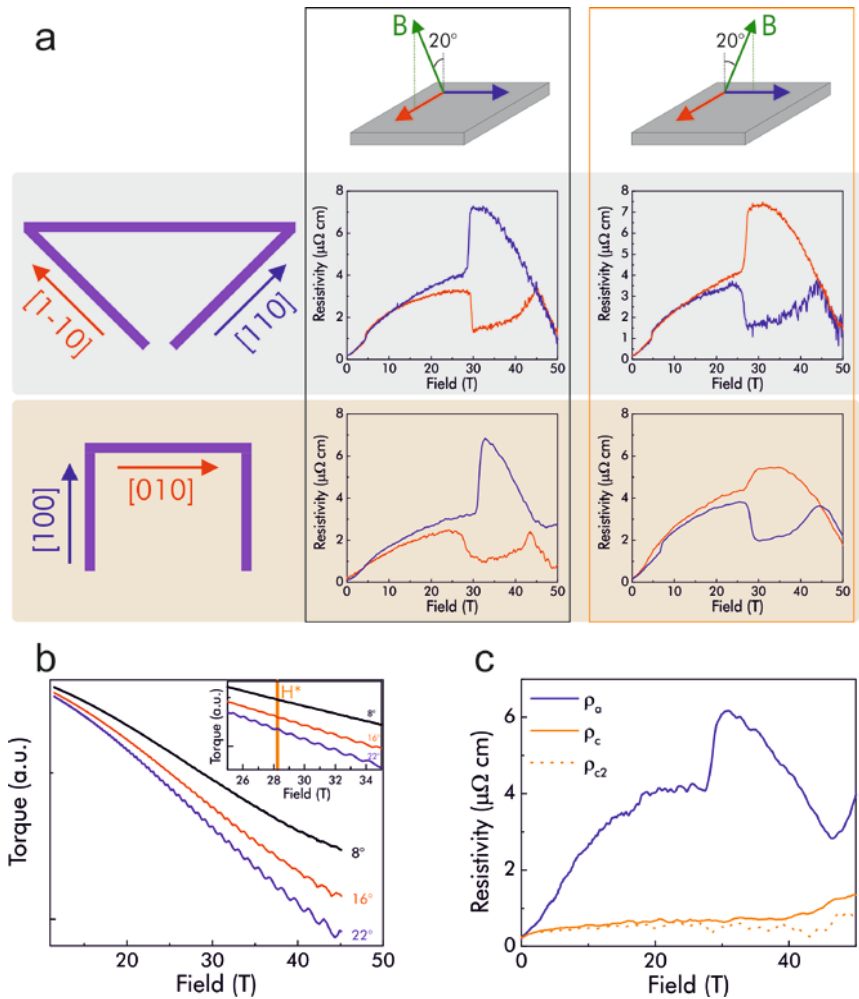

a) Magnetoresistance of the microstructures at 500mK with fields tilted 20° off the c-axis for the two structure designs shown in Figure 1. Both rows correspond to the design sketched on the left, and the colors of the resistance traces correspond to the bars of each device as indicated by the arrow colors. The high field phase is characterized by the sudden onset of a large difference of the orthogonal in-plane resistances in all studied devices. The resistance bar aligned with the in-plane field shows a dip in the resistance in all cases. When the in-plane field is tilted into the other leg, the peak and dip is interchanged (left and right column).

b) Magnetic torque at 380mK for angles 8°, 16° and 22° tilted off the c-direction into the a-direction. The torque is featureless except for de Haas-van Alphen oscillations, thus suggesting the absence of metamagnetic transitions in high fields and in particular around H*.

c) Comparison of the in-plane and out-of-plane resistivity across the transition in a microstructure at 500mK, for fields 20° off the c-direction. The high field transition is indicated by the jump in the in-plane resistivity, yet the out of plane resistivity remains featureless. This strongly suggests a highly two-dimensional electronic subsystem to be involved in the transition that does not contribute to the c-axis conduction.

## Figure 3: Temperature dependence of the anisotropy

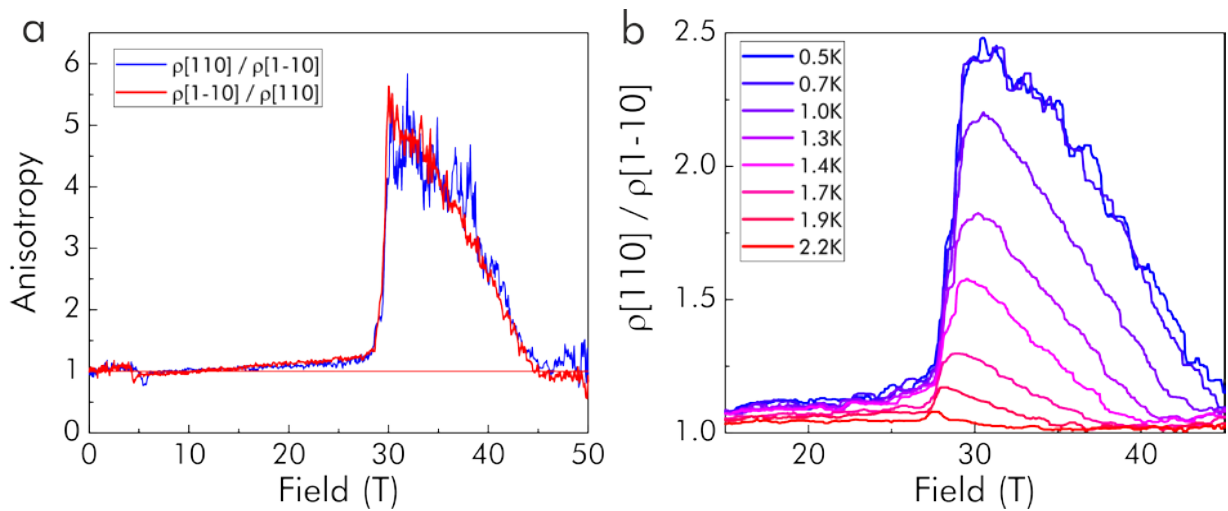

a) In-plane resistance anisotropy at 500mK, defined as the ratio between the resistance bar of high and that of low resistance. The anisotropy suddenly jumps at the transition field, and gradually shrinks to zero upon further increasing the field. When the field is tilted by 20° into the [110] direction (red trace), the anisotropy is exactly inverted compared to the case of a tilt into the [1-10] direction (blue trace).

b) Temperature dependence of the anisotropy for fields tilted 10° into the [1-10] direction. With increasing temperature, the anisotropy shrinks until it becomes unobservable at temperatures above 2.2K. H* remains temperature independent, while the anisotropy appears to continuously vanish in high fields at the Neel line $T_N(H)$.

Figure 4: Nematic character in CeRhIn$_5$

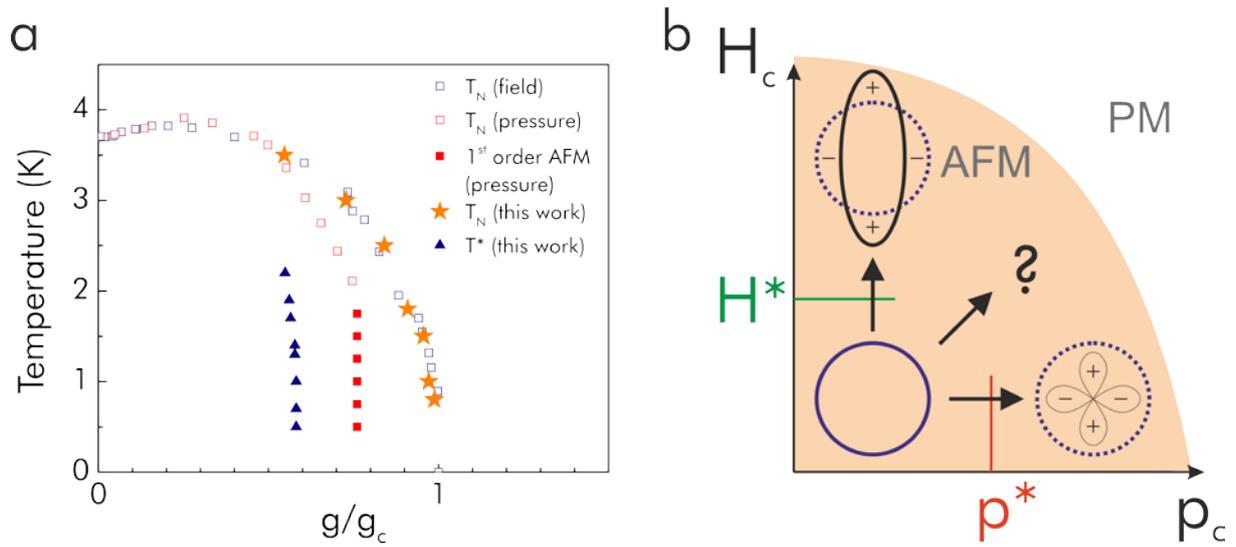

a) Generalized phase diagram comparing the suppression of the AFM order under magnetic fields and hydrostatic pressure. The dimensionless tuning parameter g/g$_c$ is defined as p/p$_c$ and H/H$_c$ in case of pressure or field respectively. The Neel line under pressure (open red symbols) leads to a vertical line of first-order phase transitions into the superconducting state (closed red symbols, both from ref. (32)). The Neel line under field (blue open symbols) (from ref.(8) ) is remarkably similar and well reproduced by our resistivity measurements (stars). Here, a vertical line of first-order phase transitions into the broken-symmetry state are observed.

b) Sketch of a unifying phase diagram. As the AFM order is suppressed, the increasing fluctuations drive a phase transition into a correlated state around the AFM critical point (p$_c$ or H$_c$). In zero field under pressure, this correlated phase appearing at p* is d-wave superconductivity. In high field at ambient pressure, a phase with an electronic nematic character results above H*. Other paths on the field-pressure plane can approach the AFM phase boundary, potentially connecting the superconducting and nematic phases.

## Acknowledgements

We thank Andy Mackenzie, Bertram Batlogg, Steve Kivelson, Christoph Geibel and Joe Thompson for stimulating discussions. L.B. is supported by DOE-BES through award DE-SC0002613.